\documentclass[aps,preprint]{revtex4}
\usepackage{amsfonts}
\usepackage{amsmath}
\usepackage{amssymb}
\usepackage{graphicx}
\usepackage{pdfpages}
\usepackage{cleveref}
\usepackage{subcaption}

\providecommand{\U}[1]{\protect\rule{.1in}{.1in}}

\begin{document}

\preprint{HEP/123-qed}
\title{Solution of Dirac equation and greybody radiation around a regular
Bardeen black hole surrounded by quintessence}
\author{Ahmad Al-Badawi}
\affiliation{Department of Physics, Al-Hussein Bin Talal University, P. O. Box: 20,
71111, Ma'an, Jordan. }
\affiliation{ahmadbadawi@ahu.edu.jo}
\author{\.{I}zzet Sakall{\i}}
\affiliation{Physics Department, Arts and Sciences Faculty, Eastern Mediterranean
University, Famagusta, North Cyprus via Mersin 10, Turkey.}
\affiliation{izzet.sakalli@emu.edu.tr}
\author{Sara Kanzi}
\affiliation{Physics Department, Arts and Sciences Faculty, Eastern Mediterranean
University, Famagusta, North Cyprus via Mersin 10, Turkey.}
\affiliation{sara.kanzi@emu.edu.tr}
\author{}
\affiliation{}

\begin{abstract}
The exact solutions of the Dirac equation that describe a massive,
non-charged particle with spin$-\frac{1}{2}$ in the curved spacetime
geometry of regular Bardeen black hole surrounded by quintessence (BBHSQ)
are investigated. We first derive the Dirac equation in the BBHSQ background
using a null tetrad in the Newman-Penrose formalism. Afterwards, we separate
the Dirac equation into ordinary differential equations for the radial and
angular parts. The angular equations are solved exactly in terms of standard
spherical harmonics. The radial part equations are transformed into
one-dimensional Schr\"{o}dinger like wave equations with effective
potentials. The effect of the quintessence on the regular Bardeen black hole
is analyzed via the physical behaviors of the effective potentials. We also
exhibit the potential graphs by changing the quintessence parameters,
magnetic monopole charge parameter, and the frequency of the particle in the
physically acceptable regions. Finally, we study the greybody factors of
bosons and fermions from the BBHSQ.
\end{abstract}

\volumeyear{}
\eid{}
\date{[]}
\received{}
\maketitle
\tableofcontents

\section{INTRODUCTION}

At astrophysics scale, observations confirm the accelerating expansion of
the universe \cite{1}. To explain the expansion, it is suggested that the
matter content in the universe has a negative pressure called dark energy 
\cite{2}. There are two kinds of negative pressure, first the cosmological
constant \cite{3,4} and the second is the so-called quintessence that causes
the acceleration of the universe \cite{5,6,7}. Quintessence is characterized
by the state equation: $p=w_{q}\rho_{q}$ where $p$ is the pressure, $%
\rho_{q} $ is the energy density, and $w_{q}$ \ is the state parameter. In
addition, the scalar fields are also hypothetical forms of dark energy where
a broad types of scalar field models have been suggested such as
quintessence \cite{8,9,10,11,12,13,14,15}, phantom models \cite%
{16,17,18,19,20}, K-essence \cite{21,22}, quintom \cite{23,24}, and so on.
Ultimately, the difference between these models is due to the magnitude of $%
w_{q}$ and for quintessence $-1\leq w_{q}\leq-\frac{1}{3}$. The quintessence
model refers to a minimally coupled scalar field with a potential which
decreases as the field increases. Quintessence is a scalar field with an
equation of state where $w_{q}$ is given by the potential energy and a
kinetic term. Hence, quintessence is dynamic and generally has a density,
and $w_{q}$ parameter varies with time. By contrast, a cosmological constant
is static, possessing a fixed energy density, and it has $w_{q}=-1.$

Usually, black holes (BHs) have singularity inside the horizon. However,
Bardeen BH (BBH) is a regular BH which does not have singularity inside the
horizon. It was first introduced by Bardeen \cite{25}. Since Bardeen
introduced his model, many other models of spherical symmetric regular BHs
were presented in the literature \cite{26,27,28,29,30,31,32,33,34,35,36}.
Later, in Ref. \cite{37,38} the authors have shown that BBH model is
explained as the gravitational field produced by a nonlinear magnetic
monopole. This explanation was extended so that it includes nonlinear
electric charge. Moreover, regular BHs surrounded by quintessence have
received major attention. Kiselev \cite{8} obtained the first analytical
solutions with spherical symmetry with quintessence surrounding the static
BHs. Later, the generalization of Kiselev solutions was obtained by
constructing their rotating counterpart \cite{39,40}. On the other hand, the
effect of quintessence on BHs have received considerable attention and their
thermodynamics has been investigated. For example, in Ref. \cite{41}\ the
thermodynamic properties of the Bardeen black hole surrounded by
quintessence (BBHSQ) was thoroughly studied.

In this paper, we consider the Dirac equation in regular BBHSQ space-time.
Recall that analytical solutions to the Dirac equation can be obtained in
several backgrounds \cite{42,43,44,45,46,47,48}. A reader is referred to see
the complete analytical solutions to the Dirac equation on de-Sitter and
anti de-Sitter space-time \cite{49,50,51,52,53}. The Dirac equation that we
consider in this study describes a massive and non-charged particle with
spin-$\frac{1}{2}$. To this end, we choose a null tetrad in order to apply
the Newman-Penrose (NP) formalism. Next, we separate the Dirac equation into
ordinary differential equations and the get coupled radial and angular
parts. The angular part equations are solved exactly in terms of standard
spherical harmonics. The radial equations are transformed into one
dimensional Schr\"{o}dinger like differential wave equations with effective
potentials. In addition, we investigate the behavior of the effective
potentials by plotting them as a function of radial distance and expose the
effect of the quintessence parameters, magnetic monopole charge parameter,
and the frequency of the particle on them. Finally, we study the outcome of
scattering a wave off a BBHSQ in terms of an absorption cross-section. The
absorption cross-section, which is a measure for the probability of an
absorption process, is directly connected to the greybody factor. Then, we
compute the greybody factor, which is nothing but the transmission
probability for an outgoing wave emitted from the event horizon of the BBHSQ
to reach the asymptotic region \cite{53v2,53v3,53v4,53v5}. Our main
motivation in the present paper paves the way to study the quasi-normal
modes associated to a field of spin-$\frac{1}{2}$ on the BBHSQ background.
Further, the given analytical expressions of the solution could be useful
for the study of the thermodynamical properties of the spinor field in same
background.

The plan of the paper is as follows. In the next section, we give a brief
discussion on the regular BBHSQ space-time. In Sec. 3, we present the Dirac
equation in BBHSQ geometry and decouple the equations into ordinary
differential equations for having the radial and angular parts. We then
obtain solutions of the angular and radial equations in Sec. 4. The
influence of the quintessence parameter is investigated through the behavior
of the effective potentials by plotting them as a function of radial
distance in the physically acceptable region. Sections 5 and 6 are devoted
to the studies of greybody factors of the BBHSQ for the spin-$0$ and spin-$%
\frac{1}{2}$ particles, respectively. Finally, we present our conclusions in
Sec. 7.

\section{Regular BBHSQ Space-time}

In this section, we shall give a brief introduction to BBHSQ which was
obtained by Kiselev \cite{8}, who assumed a spherically symmetric static
gravitational field with the following energy-momentum tensor: 
\begin{align}
T_{t}^{t}& =T_{r}^{r}=\rho _{q},  \notag \\
T_{\theta }^{\theta }& =T_{\phi }^{\phi }=-\frac{\rho _{q}}{2}\left(
3w_{q}+1\right) ,  \label{1}
\end{align}%
where $w_{q}$ is the quintessence state parameter with range $-1\leq
w_{q}\leq -1/3$ and $\rho _{q}$ is the density of the quintessence matter
given by 
\begin{equation}
\rho _{q}=-\frac{3cw_{q}}{2r^{3\left( 1+w_{q}\right) }},  \label{2}
\end{equation}%
where $c$ is the positive normalization factor ($c\geq 0$). The metric of
the regular BBHSQ\ can be expressed as \cite{41}

\begin{equation}
ds^{2}=-f\left( r\right) dt^{2}+f^{-1}(r)dr^{2}+r^{2}\left( d\theta
^{2}+\sin ^{2}\theta d\phi ^{2}\right)  \label{3}
\end{equation}%
where $f\left( r\right) $ has the following form%
\begin{equation}
f\left( r\right) =1-\frac{2Mr^{2}}{\left( r^{2}+\beta ^{2}\right) ^{3/2}}-%
\frac{c}{r^{3w_{q}+1}}.  \label{4}
\end{equation}%
in which $M$ is the mass of the BH and $\beta $\ can represent the monopole
charge of a self-gravitating magnetic field described by a nonlinear
electrodynamics source or an electric source with a field that does not
behave as the Coulomb field \cite{Coulomb}. In fact, $c$ term is related to
the density of quintessence:

\begin{equation*}
\rho _{q}=\frac{-3cw_{q}}{2r^{3\left( w_{q}+1\right) }}.
\end{equation*}%
The curvature of the metric (\ref{3}) has the form of 
\begin{equation}
R=2T_{\mu }^{\mu }=2\rho _{q}\left( 3w_{q}-1\right) ,  \label{5}
\end{equation}%
which admits a singularity at $r=0$ if $w_{q}\neq \{0,\frac{1}{3},-1\}$.
Therefore, the metric in (\ref{3}) represents a spherically symmetric
solutions for the Einstein equations describing BBHSQ with the
energy-momentum tensors given in (1). This metric satisfies all the required
limits: when $\left( c=0=\beta \right) $, we have Schwarzschild BH metric;
as $\left( c=0,\text{ }\beta \neq 0\right) $, we get Bardeen BH; and $\left(
c\neq 0,\text{ }\beta \neq 0\right) $ yields the BBHSQ.

We use the NP formalism \cite{54,55} to write and solve the Dirac equation
in the spacetime of (\ref{3}). Therefore let us define the complex null
tetrad vectors $\left( l,n,m,\overline{m}\right) $ for the metric (\ref{3})
where they satisfy the orthogonality conditions, $\left( l.n=-m\text{.}%
\overline{m}=1\right) $ as%
\begin{equation*}
l_{\mu }=dt-\frac{dr}{f\left( r\right) },
\end{equation*}%
\begin{equation*}
n_{\mu }=\frac{1}{2}f\left( r\right) dt+\frac{1}{2}dr,
\end{equation*}%
\ 
\begin{equation*}
m_{\mu }=\frac{-r}{\sqrt{2}}(d\theta +i\sin \theta d\phi ),
\end{equation*}%
\begin{equation}
\overline{m}_{\mu }=\frac{-r}{\sqrt{2}}(d\theta -i\sin \theta d\phi ),
\label{6}
\end{equation}%
and 
\begin{equation*}
l^{\mu }=f\left( r\right) dt+dr,
\end{equation*}%
\begin{equation*}
n^{\mu }=\frac{1}{2}dt-\frac{1}{2}f\left( r\right) dr,
\end{equation*}%
\ 
\begin{equation*}
m^{\mu }=\frac{1}{\sqrt{2}r}(d\theta +\frac{i}{\sin \theta }d\phi ),
\end{equation*}%
\begin{equation}
\overline{m}^{\mu }=\frac{-r}{\sqrt{2}}(d\theta -i\sin \theta d\phi ),
\label{7}
\end{equation}%
We determine the nonzero NP complex spin coefficients \cite{55} as follows 
\begin{align}
\rho & =-\frac{1}{r},\qquad \mu =\frac{1}{2r}-\frac{Mr}{\left( r^{2}+\beta
^{2}\right) ^{3/2}}-\frac{c}{2r^{3w_{q}+2}},\qquad  \notag \\
\gamma & =\frac{Mr}{2}\left[ \frac{r^{2}-2\beta ^{2}}{\left( r^{2}+\beta
^{2}\right) ^{5/2}}\right] +\frac{\left( 3w+1\right) c}{4r^{2}r^{3w}},\qquad
\alpha =-\beta _{1}=\frac{-\cot \theta }{2\sqrt{2}r}.  \label{8}
\end{align}

\section{Dirac Equation in BBHSQ}

We write the Dirac equations in the NP formalism by using the standard
notation for the spin coefficients \cite{54,55} as

\begin{equation}
\left( l^{\mu }\partial _{\mu }+\epsilon -\rho \right) F_{1}+\left( 
\overline{m}^{\mu }\partial _{\mu }+\pi -\alpha \right) F_{2}=i\mu _{0}G_{1},
\label{9}
\end{equation}%
\begin{equation}
\left( n^{\mu }\partial _{\mu }+\mu -\gamma \right) F_{2}+\left( m^{\mu
}\partial _{\mu }+\beta _{1}-\tau \right) F_{1}=i\mu _{0}G_{2},  \label{10n}
\end{equation}%
\begin{equation}
\left( l^{\mu }\partial _{\mu }+\overline{\epsilon }-\overline{\rho }\right)
G_{2}-\left( m^{\mu }\partial _{\mu }+\overline{\pi }-\overline{\alpha }%
\right) G_{1}=i\mu _{0}F_{2},  \label{11n}
\end{equation}%
\begin{equation}
\left( n^{\mu }\partial _{\mu }+\overline{\mu }-\overline{\gamma }\right)
G_{1}-\left( \overline{m}^{\mu }\partial _{\mu }+\overline{\beta }_{1}-%
\overline{\tau }\right) G_{2}=i\mu _{0}F_{1}.  \label{12n}
\end{equation}%
where $F_{1},F_{2},G_{1}$and $G_{2}$ represent the components of the wave
functions \textquotedblright Dirac spinors\textquotedblright , the mass of
the particle $\mu _{0}=\sqrt{2}\mu _{p}$ and $\epsilon ,\rho ,\pi ,\alpha
,\mu ,\gamma ,\beta _{1},\tau $ are the spin coefficients and bar over a
quantity denotes complex conjugation. We now study the Dirac equations (\ref%
{9}-\ref{12n}) in the background of metric (\ref{3}). To solve the Dirac
equations, we will consider the corresponding Compton wave of the Dirac
particle as in the form of $F=F\left( r,\theta \right) e^{i\left( kt+m\phi
\right) }$ , where $k$ is the frequency of the incoming wave and $m$ is the
azimuthal quantum number of the wave. For separable solutions, we assume 
\cite{54}, 
\begin{equation}
rF_{1}=R_{1}\left( r\right) A_{1}\left( \theta \right) \exp \left[ i\left(
kt+m\phi \right) \right] ,  \notag
\end{equation}%
\begin{equation}
F_{2}=R_{2}\left( r\right) A_{2}\left( \theta \right) \exp \left[ i\left(
kt+m\phi \right) \right] ,  \notag
\end{equation}%
\begin{equation}
G_{1}=R_{2}\left( r\right) A_{1}\left( \theta \right) \exp \left[ i\left(
kt+m\phi \right) \right] ,  \notag
\end{equation}%
\begin{equation}
rG_{2}=R_{1}\left( r\right) A_{2}\left( \theta \right) \exp \left[ i\left(
kt+m\phi \right) \right] .  \label{10}
\end{equation}%
Substituting the appropriate spin coefficients (\ref{8}) and the spinors (%
\ref{10}) into the Dirac equations (\ref{9}-\ref{12n}), we obtain the
following set of equations

\begin{equation}
A_{1}\left( \frac{d}{dr}+i\frac{k}{f}\right) R_{1}+\frac{1}{\sqrt{2}}R_{2}%
\mathbf{L}A_{2}=i\mu_{0}rR_{2}A_{1},  \notag
\end{equation}
\begin{equation}
r^{2}fA_{2}\left( \frac{d}{dr}-i\frac{k}{f}+\frac{2f+rf^{\prime}}{2rf}%
\right) R_{2}-\sqrt{2}R_{1}\mathbf{L}^{^{\dag}}A_{1}=-2i\mu_{0}rR_{1}A_{2}, 
\notag
\end{equation}
\begin{equation}
A_{2}\left( \frac{d}{dr}+i\frac{k}{f}\right) R_{1}-\frac{1}{\sqrt{2}}R_{2}%
\mathbf{L}^{^{\dag}}A_{1}=i\mu_{0}rR_{2}A_{2},  \notag
\end{equation}

\begin{equation}
r^{2}fA_{1}\left( \frac{d}{dr}-i\frac{k}{f}+\frac{2f+rf^{\prime }}{2rf}%
\right) R_{2}+\sqrt{2}R_{1}\mathbf{L}A_{2}=-2i\mu _{0}rR_{1}A_{1}.
\label{11}
\end{equation}%
where $\mathbf{L}$ and $\mathbf{L}^{^{\dag }}$ are the angular operators,
which are known as the laddering operators:

\begin{equation}
\mathbf{L}=\frac{d}{d\theta }+\frac{m}{\sin \theta }+\frac{\cot \theta }{2}%
,\qquad \mathbf{L}^{^{\dag }}=\left( \frac{d}{d\theta }-\frac{m}{\sin \theta 
}+\frac{\cot \theta }{2}\right)  \label{12}
\end{equation}%
From (\ref{11}) and \ref{12}), we get 
\begin{align}
\left( \frac{d}{dr}+i\frac{k}{f}\right) R_{1}-i\mu _{0}rR_{2}& =-\lambda
_{1}R_{2},\qquad  \notag \\
r^{2}f\left( \frac{d}{dr}-i\frac{k}{f}+\frac{2f+rf^{\prime }}{2rf}\right)
R_{2}+2i\mu _{0}rR_{1}& =\lambda _{2}R_{1},  \notag \\
\left( \frac{d}{dr}+i\frac{k}{f}\right) R_{1}-i\mu _{0}rR_{2}& =\lambda
_{3}R_{2},  \notag \\
r^{2}f\left( \frac{d}{dr}-i\frac{k}{f}+\frac{2f+rf^{\prime }}{2rf}\right)
R_{2}+2i\mu _{0}rR_{1}& =-\lambda _{4}R_{1},  \label{13}
\end{align}%
\begin{align}
\mathbf{L}A_{2}& =\lambda _{1}A_{1},\qquad \mathbf{L}^{^{\dag
}}A_{1}=\lambda _{2}A_{2},\qquad  \notag \\
\mathbf{L}^{^{\dag }}A_{1}& =\lambda _{3}A_{2},\qquad \mathbf{L}%
A_{2}=\lambda _{4}A_{1},  \label{14}
\end{align}%
The constants $\lambda _{1},$ $\lambda _{2},$ $\lambda _{3}$, and $\lambda
_{4}$ are called the separation constants. To obtain the radial and the
angular pair equations, we assume $\left( \lambda _{4}=\lambda _{1}=-\lambda
,\lambda _{2}=\lambda _{3}=\lambda \right) $, therefore (\ref{13}) and (\ref%
{14}) reduce to

\begin{equation}
\left( \frac{d}{dr}+i\frac{k}{f}\right) R_{1}=\left( \lambda+i\mu
_{0}r\right) R_{2},  \label{15}
\end{equation}

\begin{equation}
\qquad\left( \frac{d}{dr}-i\frac{k}{f}+\frac{2f+rf^{\prime}}{2rf}\right)
R_{2}=\frac{1}{r^{2}f}\left( \lambda-2i\mu_{0}r\right) R_{1},  \label{16}
\end{equation}

\begin{equation}
\mathbf{L}A_{2}=-\lambda A_{1},\qquad\mathbf{L}^{^{\dag}}A_{1}=\lambda A_{2}.
\label{17}
\end{equation}

\section{Solution of Angular and Radial Equations}

Angular equations (\ref{17}) can be rewritten as

\begin{equation}
\frac{dA_{1}}{d\theta}+\left( \frac{\cot\theta}{2}-\frac{m}{\sin\theta }%
\right) A_{1}=-\lambda A_{2},  \label{18}
\end{equation}%
\begin{equation}
\frac{dA_{2}}{d\theta}+\left( \frac{\cot\theta}{2}+\frac{m}{\sin\theta }%
\right) A_{2}=\lambda A_{1}.  \label{19}
\end{equation}
which lead to the spin-weighted spheroidal harmonics whose solution is given
in terms of standard spherical harmonics \cite{55,56,57} as

\begin{equation}
A_{1,2}=Y_{l}^{m}\left( \theta \right) ,  \label{20}
\end{equation}

with $\lambda ^{2}=\left( l+\frac{1}{2}\right) ^{2}$.

The radial equations (\ref{15}) and (\ref{16}) can be rearranged as%
\begin{equation}
\left( \frac{d}{dr}+i\frac{k}{f}\right) R_{1}=\left( \lambda +i\mu _{\ast
}r\right) R_{2},  \label{21}
\end{equation}%
\begin{equation}
r\sqrt{f}\left( \frac{d}{dr}-i\frac{k}{f}+\frac{2f+rf^{\prime }}{2rf}\right)
r\sqrt{f}R_{2}=\left( \lambda -i\mu _{\ast }r\right) R_{1},  \label{22}
\end{equation}%
where $\mu _{\ast }$ is the normalized rest mass of the spin-$\frac{1}{2}$
particle.

Our task now is to put the radial equations (\ref{21}) and (\ref{22}) in the
form of one dimensional wave equations. To this end, we follow the method
applied by Chandrasekhar's book \cite{54}. We start by making the following
transformations

\begin{equation}
P_{1}=R_{1},\qquad P_{2}=r\sqrt{f}R_{2}.  \label{23}
\end{equation}%
Hence, (\ref{21}) and (\ref{22}) transform to%
\begin{equation}
\frac{dP_{1}}{dr}+i\frac{k}{f}P_{1}=\frac{1}{r^{2}f}\left( \lambda +i\mu
_{\ast }r\right) P_{2},  \label{24}
\end{equation}%
\begin{equation}
\frac{dP_{2}}{dr}-i\frac{k}{f}P_{2}+\frac{2f+rf^{\prime }}{2rf}P_{2}=\frac{1%
}{r^{2}f}\left( \lambda -i\mu _{\ast }r\right) P_{1},  \label{25}
\end{equation}%
Assuming 
\begin{equation}
\frac{du}{dr}=\frac{1}{f},  \label{26}
\end{equation}%
then, (\ref{24}) and (\ref{25}), in terms of the new independent variable $u$%
, become%
\begin{equation}
\frac{dP_{1}}{du}+ikP_{1}=\frac{\sqrt{f}}{r}\left( \lambda +i\mu _{\ast
}r\right) P_{2},\qquad  \label{27}
\end{equation}%
\begin{equation}
\frac{dP_{2}}{du}-ikP_{2}+\frac{2f+rf^{\prime }}{2rf}P_{2}=\frac{\sqrt{f}}{r}%
\left( \lambda -i\mu _{\ast }r\right) P_{1}.  \label{28}
\end{equation}%
where 
\begin{equation}
u=r-\sqrt{M^{2}-a^{2}M^{2}-2Mr}\tan ^{-1}\left( \frac{r}{\sqrt{%
M^{2}-a^{2}M^{2}-2Mr}}\right) .  \label{29}
\end{equation}

Let us apply another transformation: 
\begin{equation}
P_{1}=\phi _{1}\exp \left[ \frac{-i}{2}\tan ^{-1}\left( \frac{\mu _{\ast }r}{%
\lambda }\right) \right] ,\qquad P_{2}=\phi _{2}\exp \left[ \frac{i}{2}\tan
^{-1}\left( \frac{\mu _{\ast }r}{\lambda }\right) \right] ,  \label{30n}
\end{equation}%
and then changing the variable $u$ into $\widehat{r}$ as $\widehat{r}=u-%
\frac{1}{2k}\tan ^{-1}\left( \frac{\mu _{0}r}{\lambda }\right) ,$ then (\ref%
{27}) and (\ref{28}) can be written in the alternative forms: 
\begin{equation}
\frac{d\phi _{1}}{d\widehat{r}}+ik\phi _{1}=W\phi _{2},  \label{30}
\end{equation}%
\begin{equation}
\frac{d\phi _{2}}{d\widehat{r}}-ik\phi _{2}=W\phi _{1},  \label{31}
\end{equation}%
where 
\begin{equation}
W=\frac{2k\sqrt{f}\left( \lambda ^{2}+\mu _{\ast }^{2}r^{2}\right) ^{3/2}}{%
2kr\left( \lambda ^{2}+\mu _{\ast }^{2}r^{2}\right) +rf\lambda \mu _{\ast }}.
\label{32}
\end{equation}

To put (\ref{30}) and (\ref{31}) into one dimensional wave equations, we
define%
\begin{equation}
2\phi _{1}=\psi _{1}+\psi _{2},\qquad 2\phi _{2}=\psi _{1}-\psi _{2}.
\label{33}
\end{equation}%
Hence, (\ref{30}) and (\ref{31}) become%
\begin{equation}
\frac{d\psi _{1}}{d\widehat{r}}-W\psi _{1}=-ik\psi _{2},  \label{34}
\end{equation}%
\begin{equation}
\frac{d\psi _{2}}{d\widehat{r}}+W\psi _{2}=-ik\psi _{1}.  \label{35}
\end{equation}
Finally, we end up with the following pair of one dimensional wave equations 
\begin{equation}
\frac{d^{2}\psi _{1}}{d\widehat{r}^{2}}+k^{2}\psi _{1}=V_{+}\psi _{1},
\label{36}
\end{equation}%
\begin{equation}
\frac{d^{2}\psi _{2}}{d\widehat{r}^{2}}+k^{2}\psi _{2}=V_{-}\psi _{2},
\label{37}
\end{equation}%
where the effective potentials can be obtained from 
\begin{equation}
V_{\pm }=W^{2}\pm \frac{dW}{d\widehat{r}}.  \label{38}
\end{equation}%
We calculate the effective potentials as\bigskip 
\begin{align}
V_{\pm }& =\frac{r^{2}B^{3}}{D^{2}}\left( 1-\frac{2Mr^{2}}{\left(
r^{2}+\beta ^{2}\right) ^{3/2}}-\frac{c}{r^{3w_{q}+1}}\right)  \notag \\
& \pm \frac{r}{D^{2}}\sqrt{B^{3}-\frac{2Mr^{2}B^{3}}{\left( r^{2}+\beta
^{2}\right) ^{3/2}}-\frac{cB^{3}}{r^{3w_{q}+1}}}\left( \left( r-M\right)
B+3r^{3}\mu _{\ast }^{2}-\frac{6r^{5}\mu _{\ast }^{2}M}{\left( r^{2}+\beta
^{2}\right) ^{3/2}}-\frac{3r^{3}\mu _{\ast }^{2}c}{r^{3w_{q}+1}}\right) 
\notag \\
& \mp \frac{r^{3}B^{5/2}}{D^{3}}\left( 1-\frac{2Mr^{2}}{\left( r^{2}+\beta
^{2}\right) ^{3/2}}-\frac{c}{r^{3w_{q}+1}}\right) ^{3/2}\left[ \left(
2rB+2r^{3}\mu _{\ast }^{2}\right) +\frac{\left( r-M\right) \lambda \mu
_{\ast }}{k}\right] ,  \label{39}
\end{align}

where 
\begin{equation}
B=\left( \lambda^{2}+\mu_{\ast}^{2}r^{2}\right) ,\qquad D=r^{2}B+\frac{%
\lambda\mu_{\ast}r^{2}}{2k}\left( 1-\frac{2Mr^{2}}{\left(
r^{2}+\beta^{2}\right) ^{3/2}}-\frac{c}{r^{3w_{q}+1}}\right) .  \label{40}
\end{equation}

Let us note that the effective potentials for the case of massless Dirac
particle (neutrino) can be obtained by setting $\mu _{\ast }=0$ in (\ref{39}%
) namely%
\begin{align}
V_{\pm }& =\lambda ^{2}\left( \frac{1}{r^{2}}-\frac{2M}{\left( r^{2}+\beta
^{2}\right) ^{3/2}}-\frac{c}{r^{3w_{q}+3}}\right) \pm \frac{\lambda \left(
r-M\right) }{r^{3}}\sqrt{\left( 1-\frac{2Mr^{2}}{\left( r^{2}+\beta
^{2}\right) ^{3/2}}-\frac{c}{r^{3w_{q}+1}}\right) }  \notag \\
& \mp \frac{2\lambda }{r^{2}}\left( 1-\frac{2Mr^{2}}{\left( r^{2}+\beta
^{2}\right) ^{3/2}}-\frac{c}{r^{3w_{q}+1}}\right) ^{3/2},  \label{41}
\end{align}

whereas substituting $c=0,$ $\beta \neq 0$ reduces to the effective
potentials for BBH

\begin{align}
V_{\pm} & =\frac{r^{2}B^{3}}{D^{2}}\left( 1-\frac{2Mr^{2}}{\left(
r^{2}+\beta^{2}\right) ^{3/2}}\right) \pm\frac{r}{D^{2}}\sqrt{B^{3}-\frac{%
2Mr^{2}B^{3}}{\left( r^{2}+\beta^{2}\right) ^{3/2}}}\left( \left( r-M\right)
B+3r^{3}\mu_{\ast}^{2}-\frac{6r^{5}\mu_{\ast}^{2}M}{\left(
r^{2}+\beta^{2}\right) ^{3/2}}\right)  \notag \\
& \mp\frac{r^{3}B^{5/2}}{D^{3}}\left( 1-\frac{2Mr^{2}}{\left(
r^{2}+\beta^{2}\right) ^{3/2}}\right) ^{3/2}\left[ \left(
2rB+2r^{3}\mu_{\ast }^{2}\right) +\frac{\left( r-M\right) \lambda\mu_{\ast}}{%
k}\right] ,  \label{42}
\end{align}

where 
\begin{equation}
B=\left( \lambda ^{2}+\mu _{\ast }^{2}r^{2}\right) ,\qquad D=r^{2}B+\frac{%
\lambda \mu _{\ast }r^{2}}{2k}\left( 1-\frac{2Mr^{2}}{\left( r^{2}+\beta
^{2}\right) ^{3/2}}\right) .  \label{43}
\end{equation}%
The complete solution of (\ref{36}) and (\ref{37}) can be obtained by the
WKB approximation method (for more details, a reader is referred to \cite%
{WKB1,WKB2}). To study the asymptotic behavior of the potentials (\ref{39}),
we can expand it up to order $O\left( \frac{1}{r}\right) ^{3}$. The
potentials (\ref{39}) for BBHSQ\ (here, we choose $w_{q}=-\frac{1}{3}$ )
behave as 
\begin{equation}
V_{\pm }\simeq \mu _{\ast }^{2}\left( 1-c\right) -\frac{2M\mu _{\ast
}^{2}\pm \mu _{\ast }c\sqrt{1-c}}{r}+\eta _{\pm }\left( \frac{1}{r}\right)
^{2}+O\left( \frac{1}{r}\right) ^{3},  \label{44}
\end{equation}%
where 
\begin{equation}
\eta _{\pm }=\left[ \lambda ^{2}\left( 1-c\right) -\frac{\lambda \mu _{\ast }%
}{k}\left( c^{2}-2c+1\right) \pm \frac{M\mu _{\ast }\left( 3c\mp 4\right) }{%
\sqrt{1-c}}\pm 5M\mu _{\ast }\sqrt{1-c}\right] ,  \label{45}
\end{equation}%
In case of neutrino, the potentials simplified to%
\begin{equation}
V_{\pm }\simeq \lambda ^{2}\left( \frac{1}{r}\right) ^{2}+O\left( \frac{1}{r}%
\right) ^{3}.  \label{46}
\end{equation}%
The first term in (\ref{44}) represents the constant value of the potential
at the asymptotic infinity. The second term corresponds to the monopole type
potential, while the third term represents the dipole type potential. As
seen from the asymptotic expansion of the potentials (\ref{44}), the effect
of adding the quintessence term $\left( \frac{c}{r^{3w_{q}+1}}\right) $ to
the lapse function $f\left( r\right) $ in the metric (\ref{3}) is observed
at all terms. From (\ref{39}), we notice that the potentials $V_{\pm }$
depend on the $c,\beta $, and $k$ parameters. We would like to remind that
the effect of the quintessence on BH has received considerable attention in
General Relativity \cite{58,59}. Therefore, it is interesting to investigate
the quintessence affect the massive and non-charged Dirac particle.

\begin{figure}[h]
\centering
\includegraphics[scale=.85]{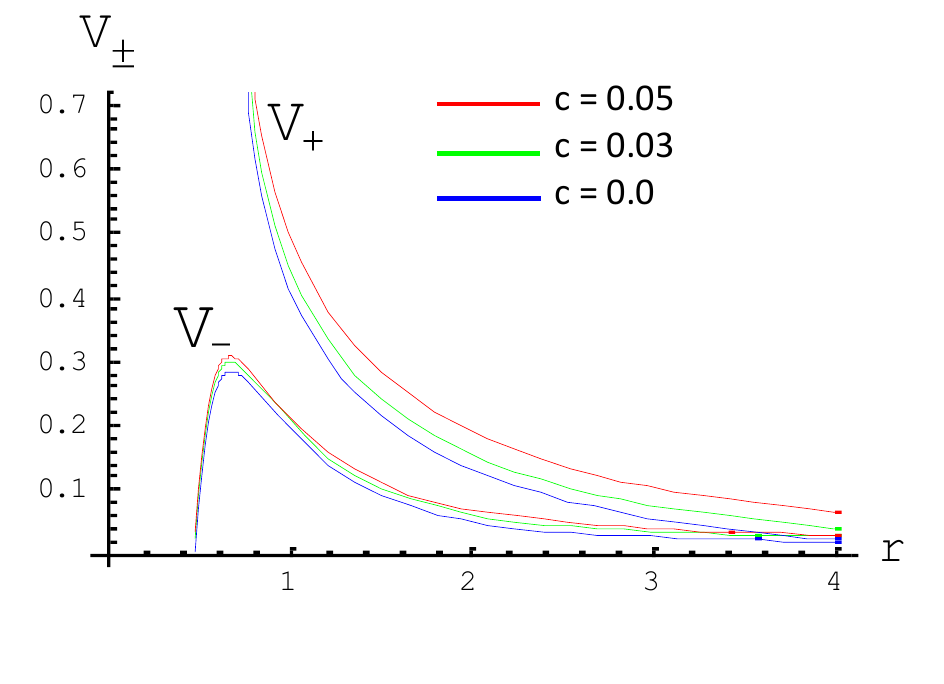}
\caption{Family of potential graphs $V_{\pm}$ for different values of $c$
with $\protect\mu^{*} = 0.12$, $k = 0.2, M = 0.5, \protect\lambda= 1, 
\protect\beta= 0.25,$ and $w_{q} = -\frac{5}{6}$.}
\label{fig1}
\end{figure}

It is obvious from (\ref{39}) that the potentials become singular when $D=0$%
. They also have local extrema when $\left( dV_{\pm }/dr\right) =0$,
however, these local extrema are very complicated algebraic equation to be
solved. To understand the physical behavior of the potentials (\ref{39}) in
the physical region and to expose the effect of the quintessence parameters $%
c$, magnetic monopole charge parameter $\beta $\thinspace\ and the frequency 
$k$ of the particle,we make two-dimensional and three-dimensional plots of
the potentials for massive particles. In all plots, we choose the fermion's
mass $\mu _{\ast }=0.12$ and the frequency $k=0.2$ such that $k>\mu _{\ast
}. $ The effective potential $V_{\pm }$ (\ref{39}) versus $r$ for different
values of $c$ is depicted in Fig. \ref{fig1}. It is observed that the
potentials have sharp peaks for all values of $c$. We notice that when the
normalization factor $c$ increases, the sharpness of the potential peaks
also increases. We deduse From Fig. \ref{fig1} that, a massive Dirac
particle in the presence of quintessence matter ($c\neq 0$) meets with a
high potential barrier, which causes a decreasing in their kinetic energies.
However, without quintessence ($c=0$) the particle encounters a low
potential barrier, which means that the Dirac particle's kinetic energy
would increase.

\begin{figure}[h]
\centering
\includegraphics[scale=.85]{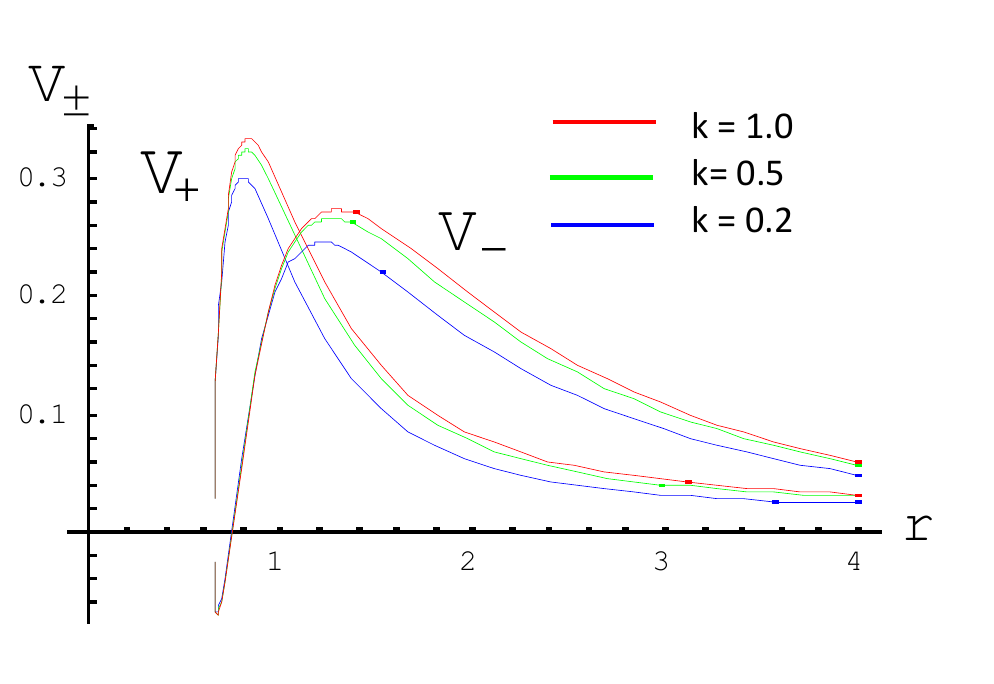}
\caption{Family of potential graphs $V_{\pm}$ for different values of
frequency $k$ with $\protect\mu^{*} = 0.12$, $c = 0.01$, $M=0.5$, $\protect%
\lambda=1$, $\protect\beta=0.25$, and $w_{q}=-1$.}
\label{fig2}
\end{figure}

In Fig. \ref{fig2}, we investigate the behavior of the effective potentials
by obtaining potential curves for some specific values of the frequency $k$
while keeping the normalization factor constant ($c=0.01$). We can see that
the potentials have peaks for all values of $k$. Again, while the frequency
increases, the potentials barrier increases, and potentials behave similarly
in the sufficiently large distances.

\begin{figure}[h]
\centering
\includegraphics[scale=.85]{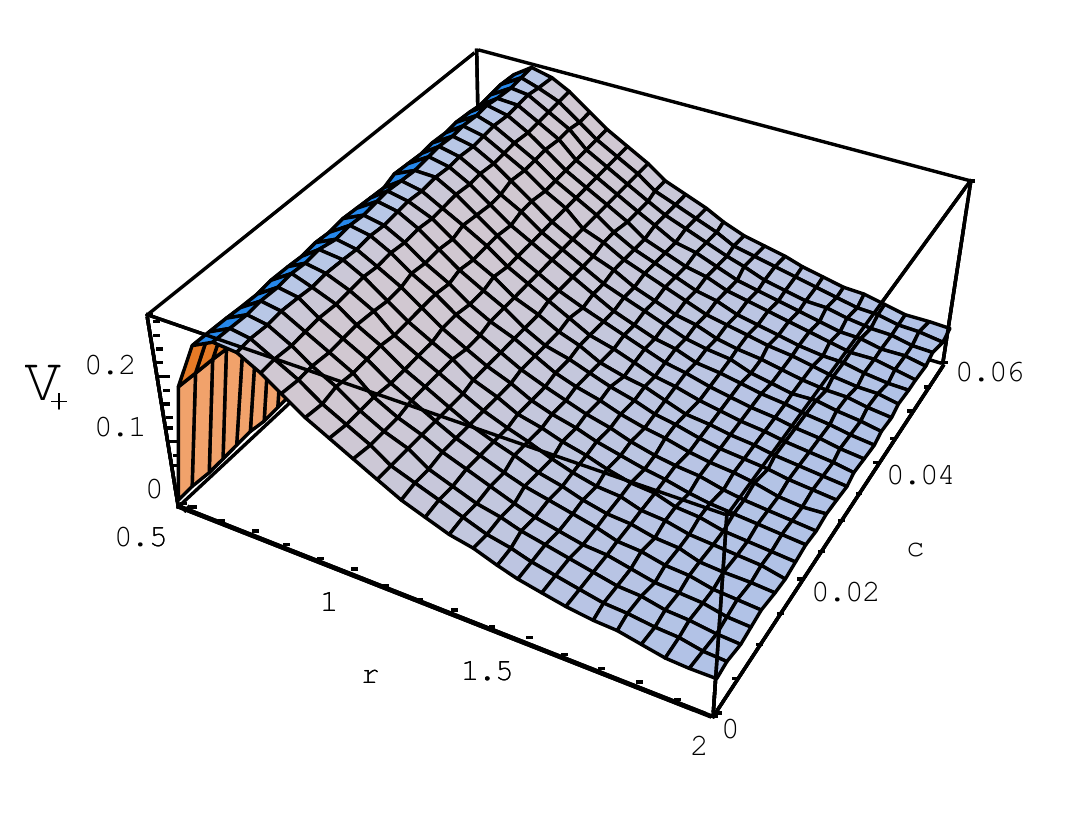}
\caption{Three-dimensional plots of potential graphs $V_{+}$ for different
values of $c$ with $\protect\mu^{*} = 0.12$, $M = 1$, $k = 0.4$, $\protect%
\lambda= 0.4$, $\protect\beta= 0.25$, and $w_{q}=-\frac{1}{3}$.}
\label{fig3}
\end{figure}

\begin{figure}[h]
\centering
\includegraphics[scale=.85]{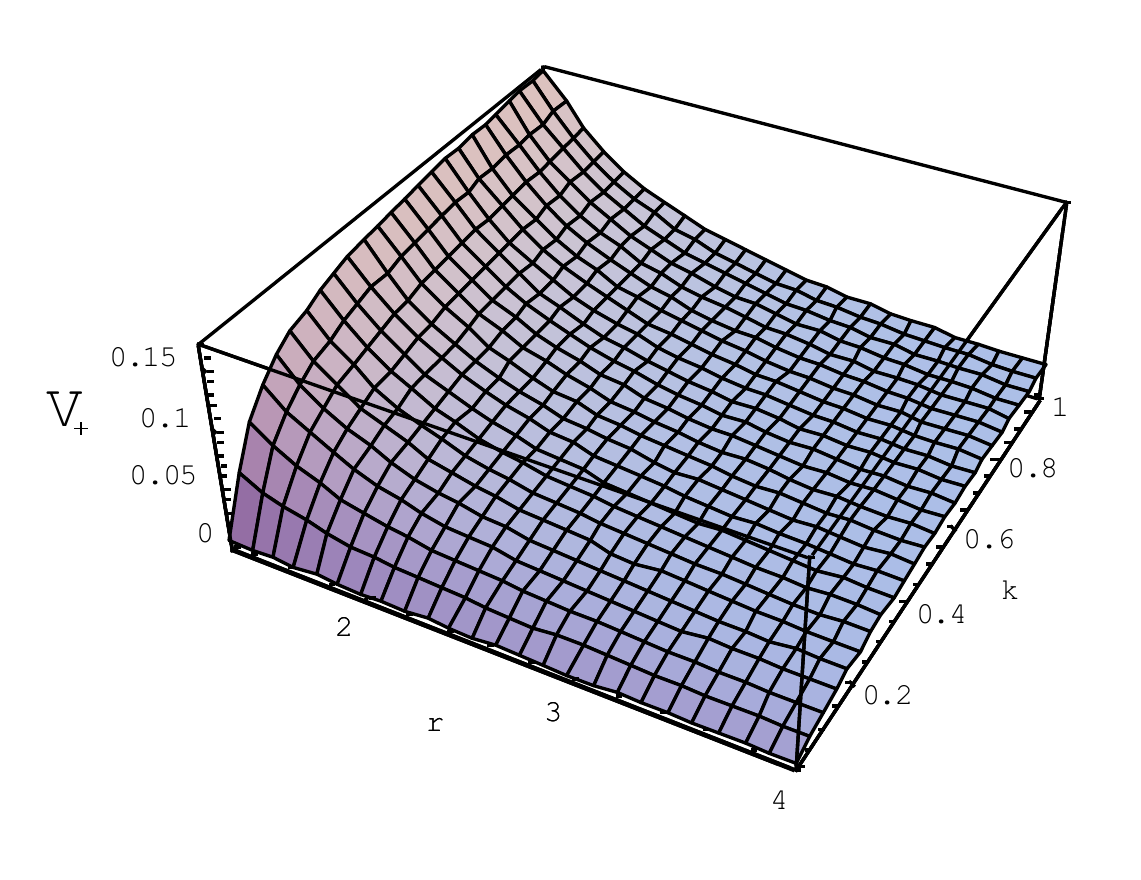}
\caption{Three-dimensional plots of potential graphs $V_{+}$ for different
values of $k$ with $\protect\mu ^{\ast }=0.12$, $M=1$, $c=0.05$, $\protect%
\lambda =0.4$, $\protect\beta =0.25$, and $w_{q}=-1$.}
\label{fig4}
\end{figure}

The effect of the quintessence term can be observed explicitly by making a
three-dimensional plot of the potential with respect to the normalization
factor $c$ and the radial distance $r$. In Fig. \ref{fig3}, we observe a
three-dimensional small peak for values of normalization factor $c$. As the
value of radial distances increases, potentials level off. Fig. \ref{fig4}
represents the three-dimensional plot of potential with respect to frequency 
$k$ and the radial distance. It is seen from Fig. \ref{fig4} that; sharp
peaks are clear for high frequencies. The effect of the monopole charge of a
self-gravitating magnetic field $\beta $ on the the potentials for the
massive charged spin-$\frac{1}{2}$ particle can be observed in Fig. \ref%
{fig5}. We deduce that; high potential barriers are observed for small
values of $\beta $ whereas for large values the potential barriers decrease.
Again, the potential levels decrease for large values of distance $r$ and
asymptote behavior is manifested. Finally, the three-dimensional plot of
potential with varying the state parameter $w_{q}$ is shown in Fig. \ref%
{fig6}. We see from Fig. \ref{fig6} that the potential levels are constant
for different values of $w_{q}$ which implies that the state parameter $%
w_{q} $ does not affect the potentials. 

\begin{figure}[h]
\centering
\includegraphics[scale=.85]{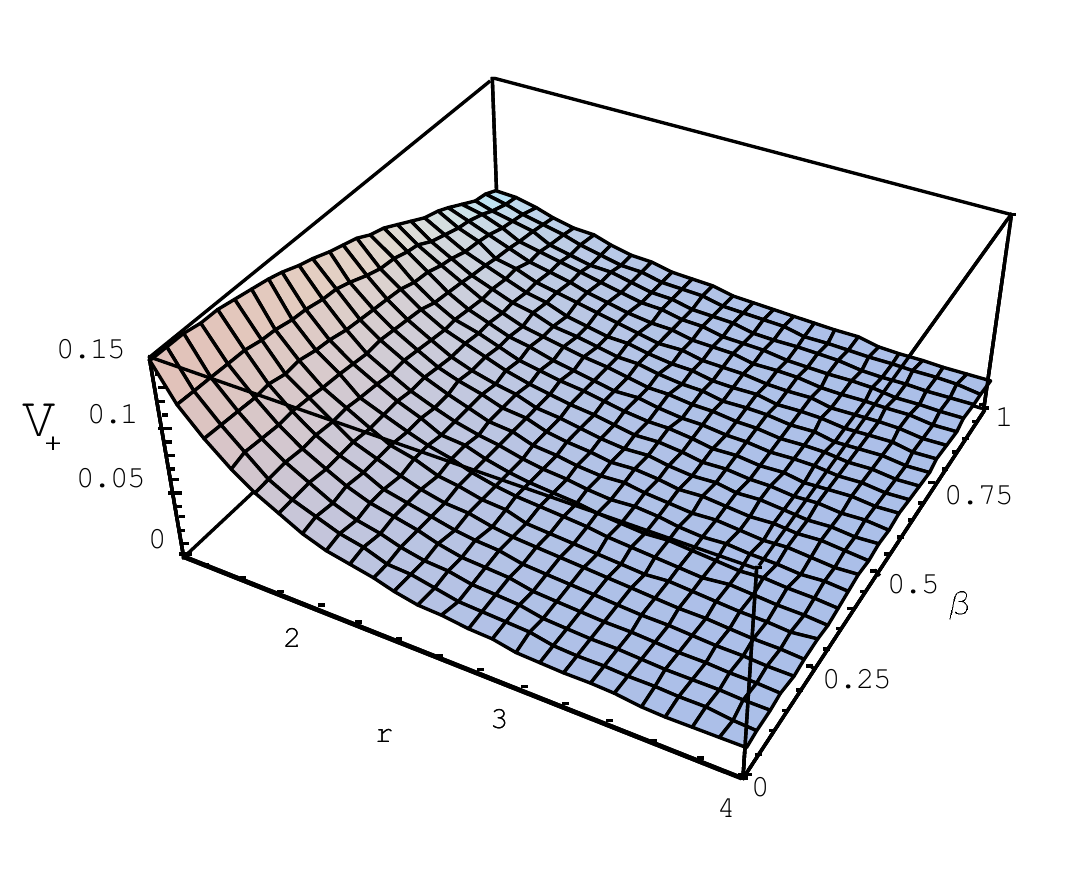}
\caption{Three-dimensional plots of potential graphs $V_{+}$ for different
values of $\protect\beta$ with $\protect\mu^{*} = 0.12$, $M = 1$, $c = 0.05$%
, $\protect\lambda= 0.4$, $k = 0.4$, and $w_{q}=-\frac{1}{3}$.}
\label{fig5}
\end{figure}

\begin{figure}[h]
\centering
\includegraphics[scale=.85]{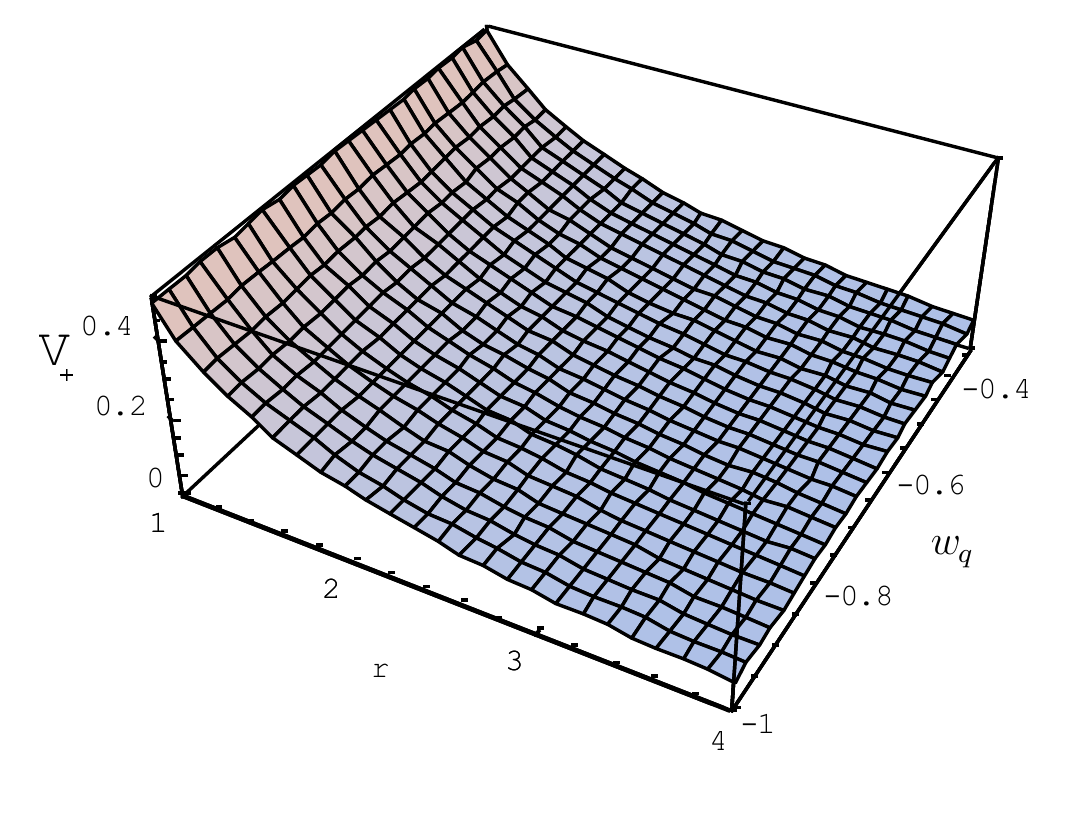}
\caption{Three-dimensional plots of potential graphs $V_{+}$ for different
values of $w_{q}$ with $\protect\mu^{*} = 0.12$, $M = 1$, $c = 0.05$, $%
\protect\lambda= 0.4$, $k = 0.4$, and $\protect\beta=0.25$.}
\label{fig6}
\end{figure}

\section{Greybody Radiation of Bosons from BBHSQ}

In this section, we evaluate the greybody factor of BBHSQ for spin$-0$
particles. For the sake of simplicity, we consider the massless Klein-Gordon
equation \cite{60}%
\begin{equation}
\square \Psi (t,r,\theta ,\phi )=0,  \label{is1}
\end{equation}

where the box symbol denotes the Laplacian operator \cite{61}:%
\begin{equation}
\square=\frac{1}{\sqrt{-g}}\partial_{\mu}\sqrt{-g}g^{\mu\nu}\partial_{\nu}.
\label{is2}
\end{equation}

By considering the metric (\ref{3}) of BBHSQ, then (\ref{is1}) reads 
\begin{equation}
-f^{-1}\partial _{t}^{2}\Psi +r^{-2}\partial _{r}\left( r^{2}f\partial
_{r}\right) \Psi +\frac{r^{-2}}{\sin \theta }\partial _{\theta }\left( \sin
\theta \partial _{\theta }\right) \Psi +\frac{r^{-2}}{\sin ^{2}\theta }%
\partial _{\phi }^{2}\Psi =0.  \label{is3}
\end{equation}

Where the scalar field can be defined as%
\begin{equation}
\Psi =p\left( r\right) A\left( \theta \right) \exp \left( -i\omega t\right)
\exp \left( im\phi \right) ,  \label{is4}
\end{equation}

here $\omega $ denotes the frequency of the wave. Substituting the the
scalar field in (\ref{is3}), one obtains%
\begin{equation}
\frac{d^{2}p(r)}{dr^{2}}+\left( f^{-1}\frac{df}{dr}+2r^{-1}\right) \frac{%
dp(r)}{dr}+\left( \omega ^{2}f^{-2}-\widehat{\lambda }r^{-2}f^{-1}\right)
p\left( r\right) =0,  \label{is5}
\end{equation}

where $\widehat{\lambda }=l\left( l+1\right) $ is the eigenvalue coming from
the physical solution of the angular equation of $A\left( \theta \right) $,
which is nothing but the standard spherical harmonics \cite{55,56,57}. By
changing the variable in a new form as $p=\frac{u}{r},$ the radial wave
equation (\ref{is5}) recasts into a one dimensional Schr\"{o}dinger like
equation as follows%
\begin{equation}
\frac{d^{2}u}{dr_{\ast }^{2}}+\left( \omega ^{2}-V_{eff}\right) u=0,
\label{is6}
\end{equation}

where the effective potential for BBHSQ is given by%
\begin{equation}
V_{eff}=f\left( \frac{\lambda}{r^{2}}+\frac{1}{r}\frac{df}{dr}\right) .
\label{is7}
\end{equation}

To evaluate the greybody factor we use \cite{54,62} 
\begin{equation}
\sigma _{l}\left( \omega \right) \geq \sec h^{2}\left( \int_{-\infty
}^{+\infty }\wp dr_{\ast }\right) ,  \label{is8}
\end{equation}

in which $r_{\ast}$ is the tortoise coordinate: $\frac{dr_{\ast}}{dr}=\frac {%
1}{f\left( r\right) },$and 
\begin{equation}
\wp=\frac{1}{2h}\sqrt{\left( \frac{dh}{dr}\right)
^{2}+(\omega^{2}-V_{eff}-h^{2})^{2}}.  \label{is9}
\end{equation}

In (\ref{is9}), $h$ is a particular positive function that satisfies the
following conditions: $h(r_{\star })>0$ and $h(-\infty )=h(\infty )=\omega $ 
\cite{63}. Here, without loss of generality, we simply set it as $h=\omega $ 
\cite{62,63}, resulting in that the integration of (\ref{is8}) becomes 
\begin{equation}
\sigma _{l}\left( \omega \right) \geq \sec h^{2}\frac{1}{2\omega }%
\int_{r_{h}}^{+\infty }\left( \frac{\lambda }{r^{2}}+\frac{1}{r}\frac{df}{dr}%
\right) dr.  \label{is92}
\end{equation}

After making a straightforward calculation, one finds 
\begin{equation}
\sigma_{l}\left( \omega\right) \geq\sec h^{2}\left( \frac{1}{2\omega }\left[ 
\frac{l\left( l+1\right) }{r_{h}}+\frac{c\left( 3w_{q}+1\right) }{\left(
3w_{q}+2\right) r_{h}^{3w_{q}+2}}-\frac{2M}{\beta^{2}}+\frac {2M}{\beta^{2}%
\sqrt{1+\frac{\beta^{2}}{r_{h}^{2}}}}+\frac{2M}{r_{h}^{2}\left( 1+\frac{%
\beta^{2}}{r_{h}^{2}}\right) ^{3/2}}\right] \right) ,  \label{is10}
\end{equation}

\bigskip where $r_{h}$\ represents the event horizon.

\begin{figure}[h]
\centering
\includegraphics[scale=.5]{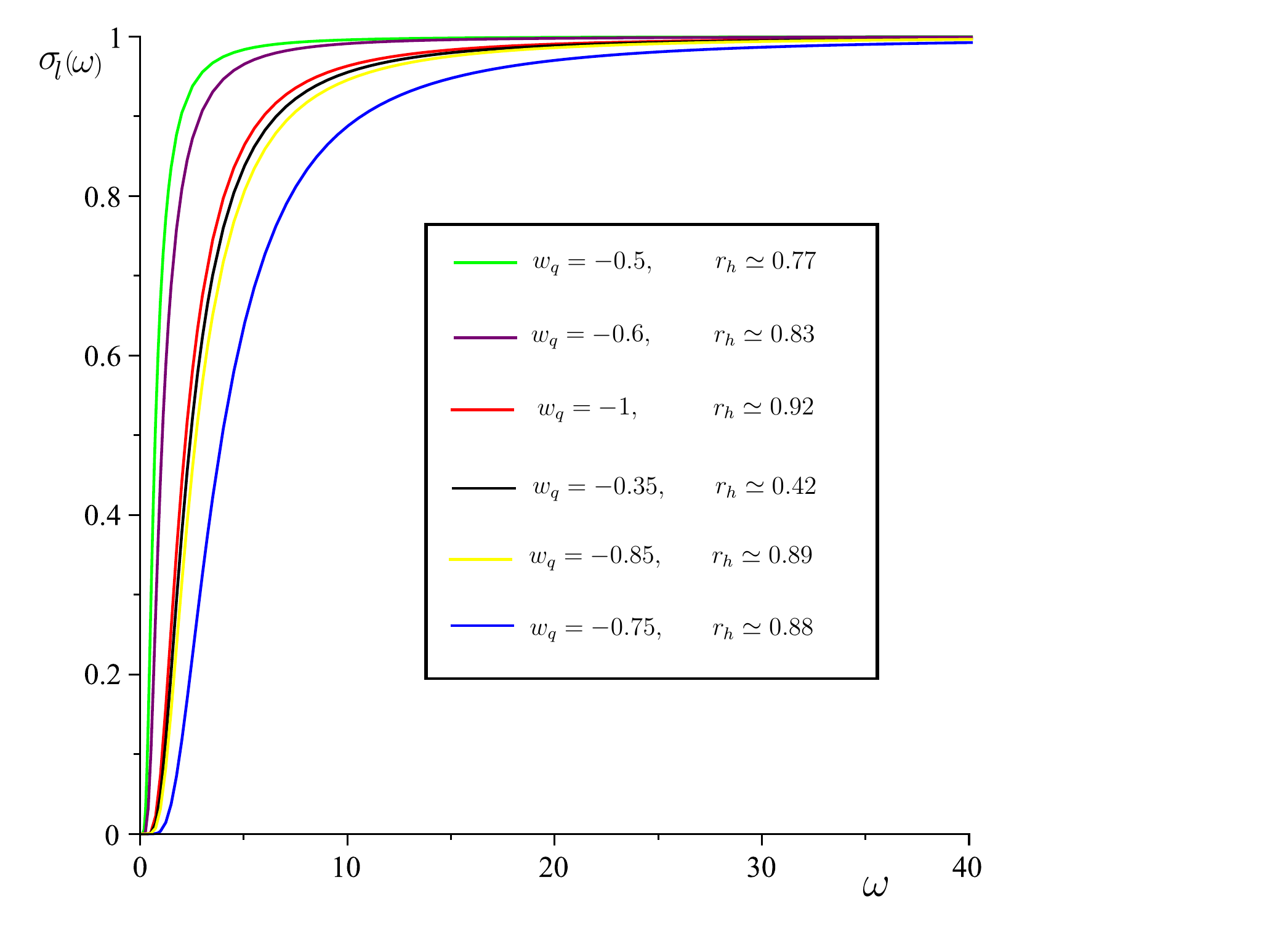}
\caption{$\protect\sigma_{l}\left( \protect\omega\right) $ versus $\protect%
\omega$ graph. The plots are governed by Eq. (\protect\ref{is10}). For
different $w_{q}$ values, the corresponding event horizons (i.e, $f(r_{h})=0$%
) are illustrated. The physical parameters for this plot are chosen as $%
M=l=c=1$, and $\protect\beta=2$.}
\label{fig7}
\end{figure}

We depict the greybody factors of the BBHSQ in Fig. \ref{fig7}. As seen from
Fig. \ref{fig7}, the values of $w_{q}$ and event horizon ($r_{h}$) are
linearly proportional to each other. It is obvious that greybody radiation
strictly depends also on the state parameter $w_{q}$. According to the
information we obtained from the graph, a similar radiation emission occurs
around critical $w_{q}$ values ($-\frac{1}{3}$ and $-1$). However, while $%
w_{q}$ value moves away from those critical values, then radiation may
decrease or increase depending on $w_{q}$.

\section{Greybody Radiation of Fermions from BBHSQ}

In this section, we shall derive the fermionic greybody factors of the
neutrinos emitted from BBHSQ. To this end, we consider the case of $w_{q}=-%
\frac{1}{3}$ in order to obtain analytical results from Eq. (\ref{is8}). In
the case of $w_{q}=-\frac{1}{3}$, the potentials (\ref{41}) can be rewritten
as 
\begin{equation}
V_{\pm }=\frac{\lambda }{r^{2}}f\pm \frac{\lambda \left( r-M\right) }{r^{3}}%
\sqrt{f}\mp \frac{2\lambda }{r^{2}}f^{3/2},  \label{is11}
\end{equation}

in which

\begin{equation}
f=1-\frac{2Mr^{2}}{\left( r^{2}+\beta ^{2}\right) ^{3/2}}-c.  \label{is12}
\end{equation}

Following the procedure described in the previous section [see Eqs. (\ref%
{is8})-(\ref{is11})], one can get%
\begin{equation}
\sigma _{l}^{\pm }\left( \omega \right) \geq \sec h^{2}\left( \frac{1}{%
2\omega }\int_{r_{h}}^{+\infty }\left[ \frac{\lambda }{r^{2}}\pm \left( 
\frac{\lambda }{r^{2}}-\frac{\lambda M}{r^{3}}\right) \frac{1}{\sqrt{f}}\mp 
\frac{2\lambda }{r^{2}}\sqrt{f}\right] dr\right) ,  \label{is13}
\end{equation}

in which $\sigma _{l}^{+}\left( \omega \right) $ and $\sigma _{l}^{-}\left(
\omega \right) $ stand for the greybody factors of the spin-up and spin-down
fermions, respectively. After performing some tedious computations, the
greybody factors of the fermions can be obtained as follows%
\begin{align}
\sigma _{l}^{\pm }\left( \omega \right) & \geq \sec h^{2}\left\{ \frac{1}{%
2\omega }\left[ \frac{\lambda }{r}\pm \left( \frac{\lambda }{r_{h}\sqrt{1-c}}%
+\frac{M\lambda }{2\left( 1-c\right) ^{3/2}r_{h}^{2}}+\frac{M^{2}\lambda }{%
2\left( 1-c\right) ^{5/2}r_{h}^{3}}+\right. \right. \right.  \notag \\
& \left. \frac{-3\lambda M\beta ^{2}(1-c)^{2}+5M^{3}\lambda }{%
8r_{h}^{4}(1-c)^{3}\sqrt{1-c}}+\frac{-9M^{2}\beta ^{2}\lambda (1-c)^{2}+%
\frac{35}{4}M^{4}\lambda }{10r_{h}^{5}\left( 1-c\right) ^{4}\sqrt{1-c}}-%
\frac{\lambda M}{2\sqrt{1-c}r_{h}^{2}}-\frac{\lambda M^{2}}{3\left(
1-c\right) ^{3/2}r_{h}^{3}}-\right.  \notag \\
& \left. \left. \frac{3\lambda M^{3}}{8r_{h}^{4}\left( 1-c\right) ^{5/2}}-%
\frac{-3\lambda M^{2}\beta ^{2}(1-c)^{2}+5\lambda M^{4}}{10r_{h}^{5}\left(
1-c\right) ^{3}\sqrt{1-c}}-\frac{-9\lambda M^{3}\beta ^{2}(1-c)^{2}+\frac{35%
}{4}\lambda M^{5}}{12r_{h}^{6}\left( 1-c\right) ^{4}\sqrt{1-c}}\right.
\right)  \notag \\
& \mp \left( \frac{2\lambda \sqrt{1-c}}{r_{h}}-\frac{\lambda M}{\sqrt{1-c}%
r_{h}^{2}}-\frac{\lambda M^{2}}{3\left( 1-c\right) ^{3/2}r_{h}^{2}}+\frac{%
\lambda \sqrt{1-c}\left( 3M\beta ^{2}(1-c)^{2}-M^{3}\right) }{%
4r_{h}^{4}\left( 1-c\right) ^{3}}\right.  \notag \\
& \left. \left. \left. +\frac{\lambda \sqrt{1-c}\left( 12M^{2}\beta
^{2}(1-c)^{2}-5M^{4}\right) }{20r_{h}^{5}\left( 1-c\right) ^{4}}\right) %
\right] \right\} \text{ \ \ \ \ \ \ }\left( \text{ for }0\leq c<1\right) ,
\label{is14}
\end{align}

\begin{figure}[h]
\begin{subfigure}{.5\textwidth}
  \centering
  \includegraphics[width=.8\linewidth]{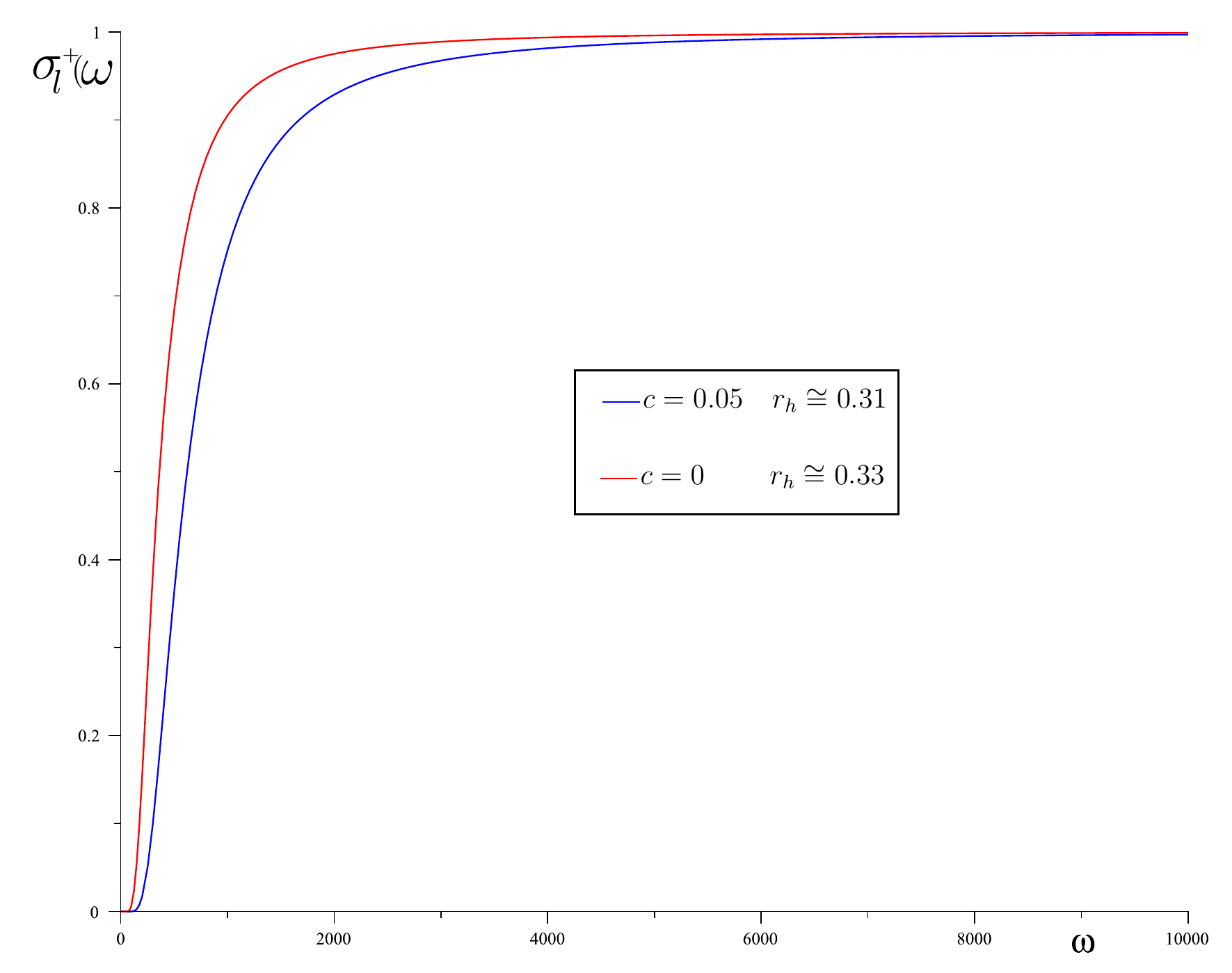}
  \caption{For $c\approx0$ values.}
  \label{fig8a}
\end{subfigure}%
\begin{subfigure}{.5\textwidth}
  \centering
  \includegraphics[width=.8\linewidth]{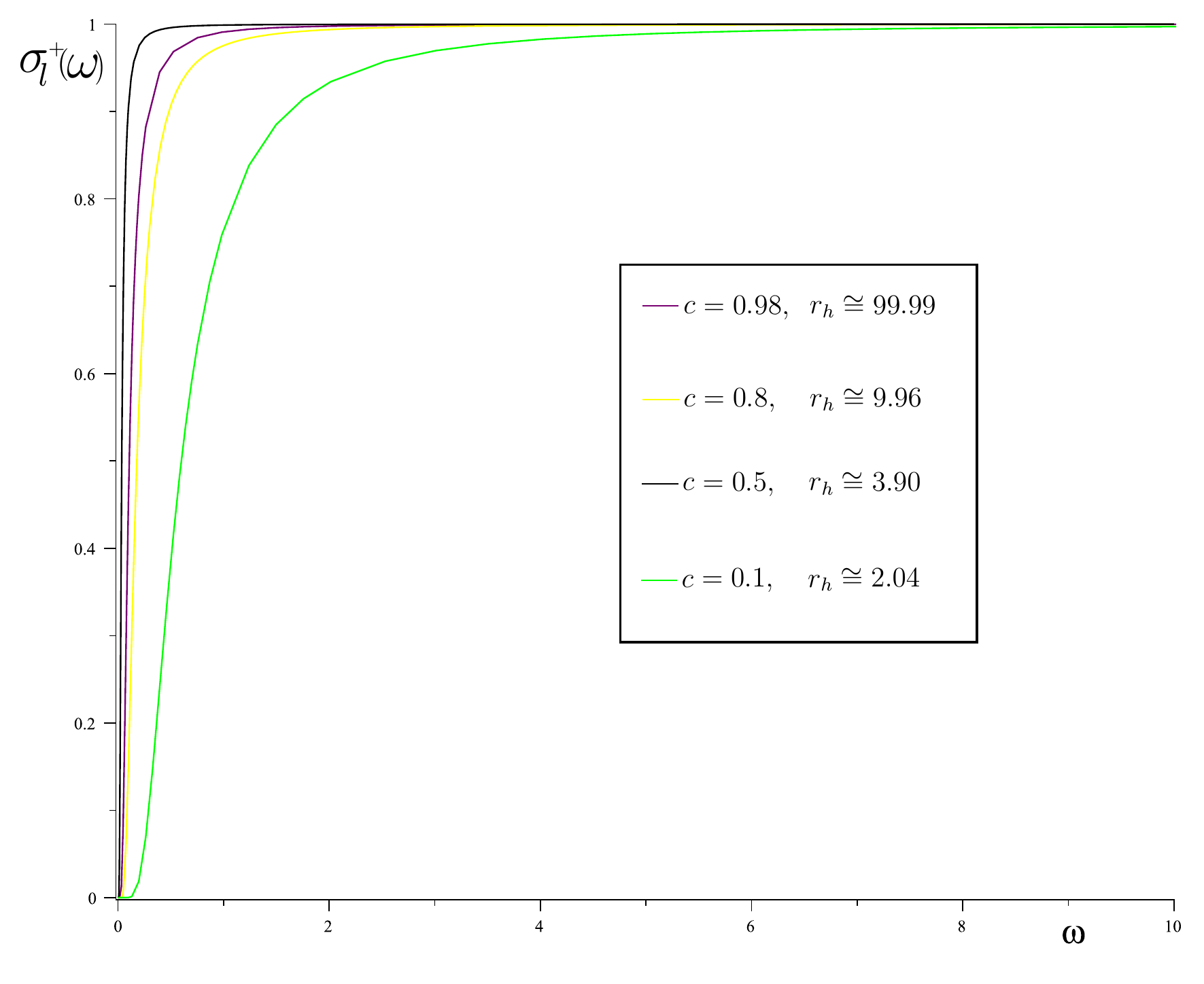}
  \caption{For $c>0$ values.}
  \label{fig8b}
\end{subfigure}
\caption{$\protect\sigma _{l}^{+}\left( \protect\omega\right) $ versus $%
\protect\omega$ graph for the case of $w_{q}=-\frac{1}{3}$. The plots are
governed by Eq. (\protect\ref{is14}). For different $c$ values, the
corresponding event horizons (i.e, $f(r_{h})=0$) are illustrated. The
physical parameters for these plots are chosen as $M=l=1$, and $\protect\beta%
=0.5$.}
\label{fig8}
\end{figure}

\begin{figure}[h]
\begin{subfigure}{.5\textwidth}
  \centering
  \includegraphics[width=.8\linewidth]{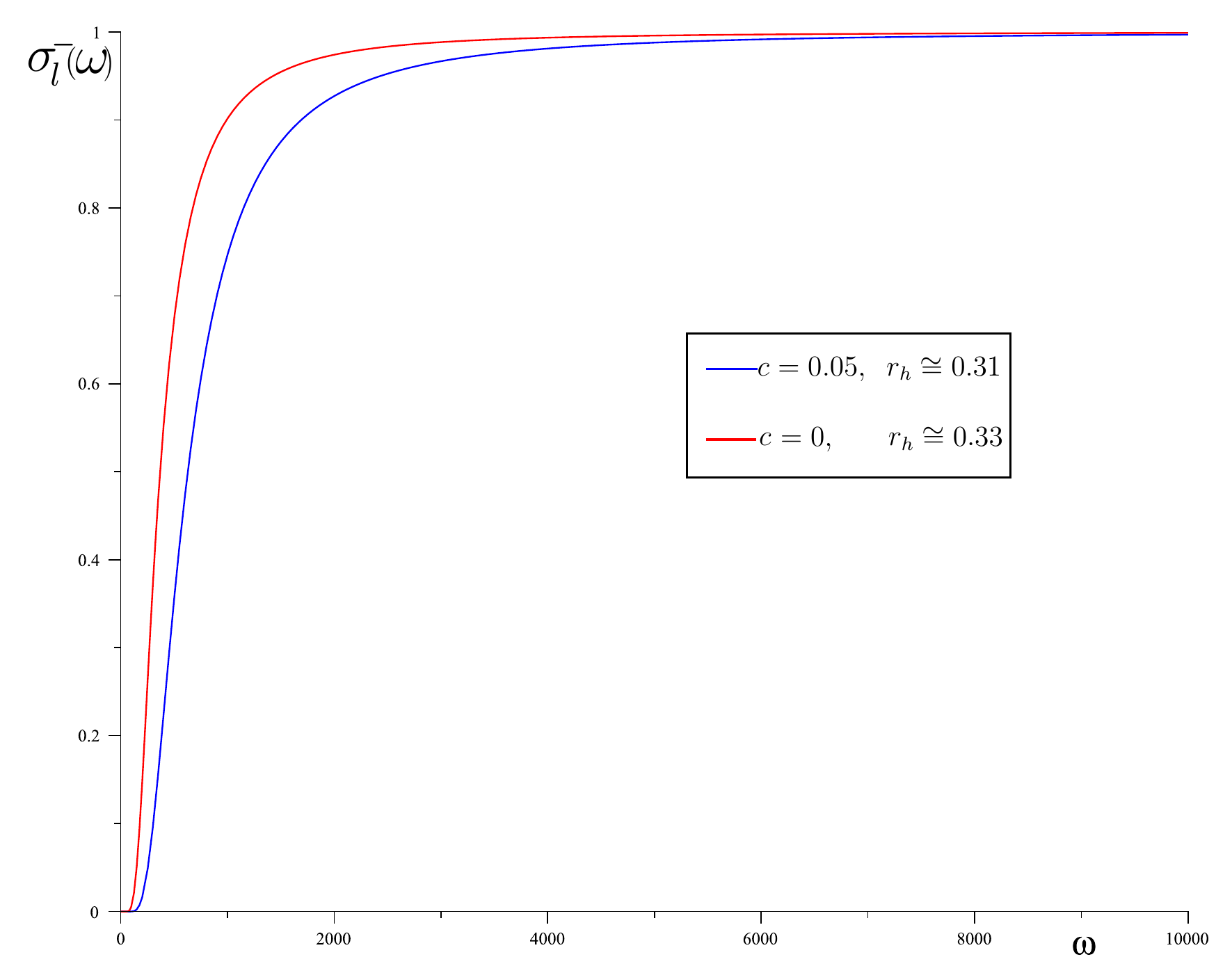}
  \caption{For $c\approx0$ values.}
  \label{fig9a}
\end{subfigure}%
\begin{subfigure}{.5\textwidth}
  \centering
  \includegraphics[width=.8\linewidth]{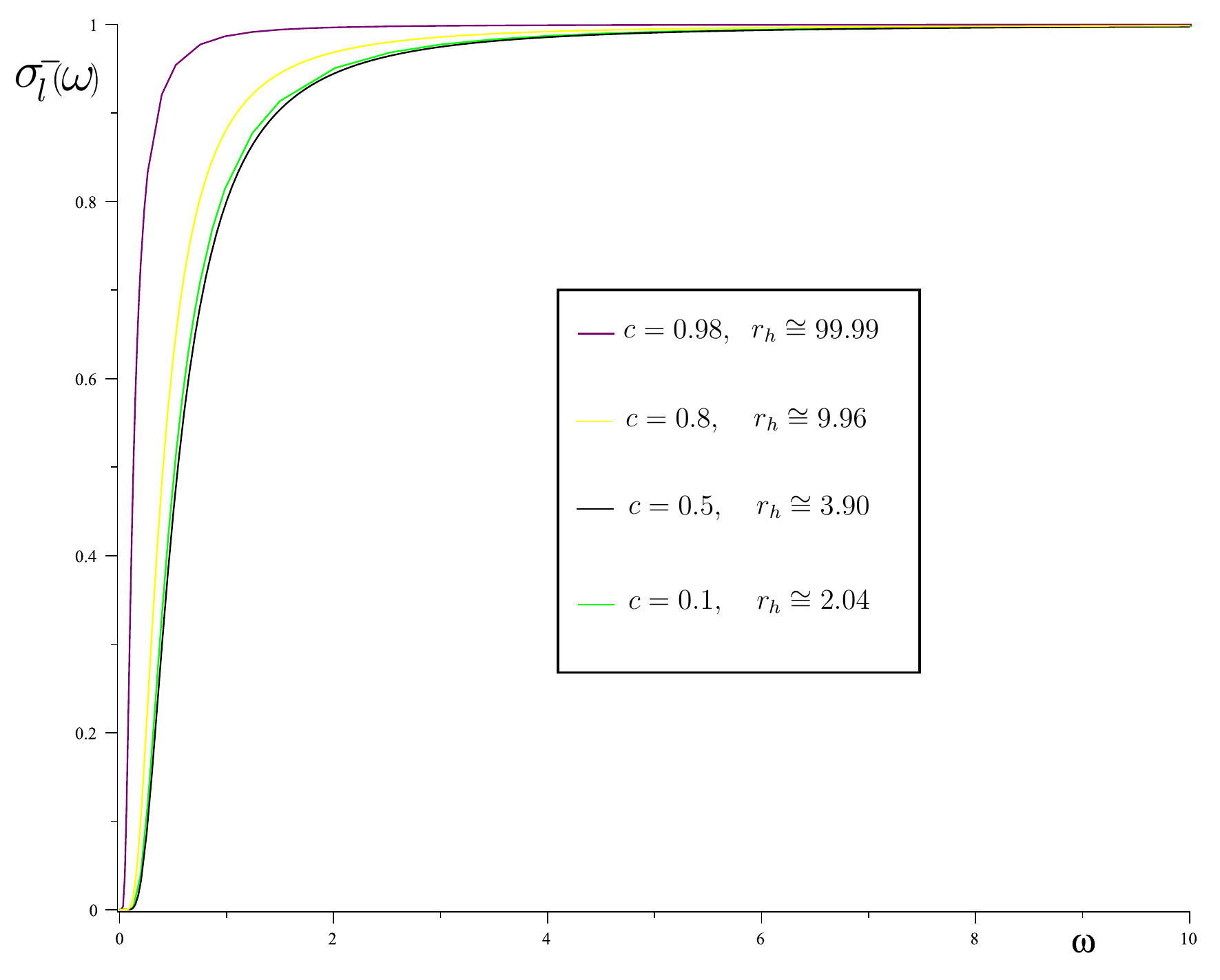}
  \caption{For $c>0$ values.}
  \label{fig9b}
\end{subfigure}
\caption{$\protect\sigma _{l}^{-}\left( \protect\omega \right) $ versus $%
\protect\omega $ graph for the case of $w_{q}=-\frac{1}{3}$. The plots are
governed by Eq. (\protect\ref{is14}). For different $c$ values, the
corresponding event horizons (i.e, $f(r_{h})=0$) are illustrated. The
physical parameters for these plots are chosen as $M=l=1$, and $\protect%
\beta =0.5$.}
\label{fig9}
\end{figure}
As it can been seen from Figs. (\ref{fig8}) and (\ref{fig9}), the greybody
factors of spin-up and spin-down fermions exhibit almost the same behaviors
as $c$ changes.

\section{Conclusion}

In this paper, we have investigated the exact solutions of the Dirac
equations that describe a massive, non-charged particle with spin-$\frac{1}{2%
}$ in the curved space-time geometry of BBHSQ, using NP\ (null tetrad)
formalism. By employing an axially symmetric ansatz for the Dirac spinors,
we decouple equations into angular and radial parts. The angular equation
leads to the spin-weighted spheroidal harmonics with eigenvalue $\lambda
^{2}=\left( l+\frac{1}{2}\right) ^{2}$. The radial equations were reduced to
pair of one-dimensional Schrodinger-like wave equations with effective
potentials for the Dirac particle. We then studied the potentials by
plotting them as a function of radial distance. Thus, the effect of the
quintessence term on the BBH are unfolded. We revealed that potentials
barriers having quintessence matter become more higher than the potentials
without the quintessence. We also showed that, as the frequency increases,
potentials levels increase as well. However, as the magnetic monopole charge
parameter $\beta $ increases, the potential levels decrease whereas the
potentials do not change for varying the state parameter $w_{q}$.
Remarkably, we depicted how the greybody factors of bosons and fermions
vary with the quintessence state parameters $w_{q}$ (see Fig. \ref{fig7})
and $c$ (see Figs. \ref{fig8} and \ref{fig9}), respectively.

In future work, we will extend our analysis to the Dirac equation of charged
massive fermionic waves propagating in the rotating geometry of the BBHSQ.
In this way, we plan to analyze the effect of quintessence on the stationary
spacetimes using fermions. For this purpose, we shall also consider the
Janis-Newman algorithm \cite{64} for the static BBHSQ (\ref{3}).

\end{document}